\documentclass[twocolumn]{webofc}

\usepackage[varg]{txfonts}   
\usepackage{hyperref}
\usepackage{url}
\hypersetup{colorlinks=true,citecolor=blue,urlcolor=blue,linkcolor=blue}
\begin{document}
\title{Transitions To Door-way States And Nuclear Responses Against 2-body External Fields}
%
%

\author{\firstname{Futoshi} \lastname{Minato}\inst{1,2,3}\fnsep\thanks{\email{minato.futoshi.009@m.kyushu-u.ac.jp}}
}

\institute{
Department of Physics, Kyushu University, Fukuoka 819-0395, Japan
\and
RIKEN Nishina Center for Accelerator-Based Science, Wako, Saitama 351-0198, Japan
\and
Nuclear Data Center, Japan Atomic Energy Agency, Tokai, Ibaraki 319-1195, Japan
}

\abstract{
Nuclear microscopic structural models that treat two-body effective interactions self-consistently becomes available, one of which is second-random-phase-approximation (SRPA).
SRPA can be used to study evolutions from 1 particle-1 hole (1p1h) to 2 particle-2 hole (2p2h) states from different point of view from reaction models. We studied nuclear excitations created by 1-body and 2-body external fields and discuss transitions between 1p1h and 2p2h states obtained by the SRPA approach.
}
\maketitle
\section{Introduction}
\label{intro}
Bohr's hypothesis for nuclear compound states has succeeded in describing low-energy particle emissions.
In the shell model picture, this idea works well if 1p1h states created by an external field transfer instantly to many particle-many hole ($m$p$m$h) states, which are called compound or thermal equilibrium states.
However, if the initially excited particle has a large energy and its mean free path is longer than a nuclear size (typically around $E^{*} >$ a few 10~MeV), the particle can escape from the nucleus before the compound state is formed. 
This process is the so-called particle emissions from pre-equilibrium states.
In theory, compound and pre-equilibrium states have been discussed in various nuclear reaction models~\cite{Carlson2014}.
Multi-step direct reaction model~\cite{FKK, Tamura1982, Nishioka1988} and multi-step compound reaction models~\cite{Agassi1975, Nishioka1986} gave important insights on transition from 1 particle-0 hole states, the initial channel of nucleon-nucleus reactions, to 2 particle-1 hole and more complicated states.
For practical applications, exciton models~\cite{Griffin1966} and Hauser-Feshbach models~\cite{HauserFeshbach} are known as an effective tool to describe particle emissions from pre-equilibrium and compound states, respectively, and these are widely applied not only in analyses of fundamental nuclear physics but also nuclear data evaluations.
\par
Recently, nuclear microscopic structural models that treat two-body effective interactions self-consistently becomes available with increasing available computer resources.
In particular, second-random-phase-approximation (SRPA)~\cite{Providencia1965}, which describes nuclear excitations up to 2p2h states, has been applied to study spreading widths of giant resonances and Gamow-Teller transitions~\cite{Gambacurta2015, Gambacurta2020, Yang2021, Gambacurta2022, Yang2023}.
We notice that SRPA can be also used to study evolutions from 1p1h to 2p2h states from different point of view from the aforementioned reaction models.
In this paper, we present our recent findings that provide us how 1p1h states created by an external field evolve to 2p2h states in the framework of SRPA approach.
Here, we should mention that there is a phonon-phonon coupling model~\cite{Severyukhin2004,Gambacurta2016,Arsenyev2023} that describes higher-order effect by mapping 2p2h states with multi-phonon states.
\par
In addition, we also discuss nuclear excitations induced by 2-body external fields.
External fields are usually represented by 1-body operators exciting 1p1h states in nuclei. 
However, nuclei can be excited into 2p2h states directly as an initial configuration for some cases through 2-body operators.
Typical example is a meson-exchange current (MEC) that have 2-body components mediated by meson currents.
We recently studies the effect of MEC on particle emissions following negative muon captures on nuclei~\cite{Minato2023}.
As already pointed out in Ref.~\cite{Lifshitz1980}, we confirmed that the MEC is importance to reproduce high-energy proton emissions by introducing a phenomenological MEC effect.
Although there are a number of studies on nuclear excitations for 1-body external fields, the transition probabilities of 2-body external fields have not been studied well and its understanding is inadequate.
Microscopic understanding of nuclear excitations induced by 2-body external fields in nuclei is thus highly demanded for muon capture reactions.
\par
The contents of this paper is as follows. 
Sect.~\ref{sect:theo} explains the theoretical framework that we used. Sect.~\ref{sect:result} discusses our result, and Sect.~\ref{sect:summary} summarizes the discussion in this paper.
\section{Theoretical Framework}
\label{sect:theo}
SRPA is a useful tool to study 2p2h excitations for nuclei from light to heavy nuclei.
We developed a new computational code for SRPA.
Here, we demonstrate its formalism briefly.
In the SRPA, the phonon creation operator $Q_{\nu}^{\dagger}$ for resonance $\nu$ is defined as~\cite{Papa}
\begin{equation}
\begin{split}
Q^{\dagger}_{\nu}
&=\sum_{mi} X_{mi}^{(\nu)} O_{mi}^{\dagger}
+\sum_{m<n,i<j} \mathcal{X}_{mnij}^{(\nu)} O_{mnij}^{\dagger}\\
&-\sum_{mi} Y_{mi}^{(\nu)} O_{im}^{\dagger}
-\sum_{m<n,i<j} \mathcal{Y}_{mnij}^{(\nu)} O_{ijmn}^{\dagger}.
\end{split}
\end{equation}
Indices $m(n)$ and $i(j)$ indicate particle and hole states of single-particle states, which are calculated by Skyrme-Hartree-Fock method with the SkM$^{*}$ interaction~\cite{Bartel1982} and box size $16$ fm.
The operators $O_{mi}^{\dagger}$ and $O_{mnij}^{\dagger}$ create 1p1h and 2p2h states, respectively, and its detailed form can be found in Ref.~\cite{Papa}.
Neglecting the terms of $\mathcal{X}_{mnij}^{(\nu)}$ and $\mathcal{Y}_{mnij}^{(\nu)}$, the formalism corresponds to the standard 1p1h RPA.
The coefficients of $X_{mi}^{(\nu)}, Y_{mi}^{(\nu)}, \mathcal{X}_{mnij}^{(\nu)}, \mathcal{Y}_{mnij}^{(\nu)}$, and SRPA phonon energy $E_{\nu}$ are determined by solving the SRPA equation.
\par
The transition strength of the 1-body external field $O_{J}$ without a spin operator is represented by
\begin{equation}
\left|\langle \nu || O_{J} || 0 \rangle\right|^{2}
=\left|\sum_{mi} \Big(X_{mi}^{(\nu)}+(-1)^{J}Y_{mi}^{(\nu)}\Big)O_{mi}^{(J)}\right|^{2}
\label{eq:1tran}
\end{equation}
while the transition strength of the double phonon operator $T_{J}\equiv\left[O_{L}P_{L}\right]^{(J)}$ is given by
\begin{equation}
\begin{split}
\left|\langle \nu||T_{J}||0\rangle\right|^{2}
&=\left|\sum_{mi}\Big(X_{mi}^{(\nu)}+(-1)^{J}Y_{mi}^{(\nu)}\Big)T_{mi}^{(J)}\right.\\
&\left.+\sum_{mnij}\sum_{J_{p}J_{h}}\Big(\mathcal{X}_{mnij}^{(\nu)}+\mathcal{Y}_{mnij}^{(\nu)}\Big)T_{mnij}^{(J_{p}J_{h}J)}\right|^{2}
\end{split}
\label{eq:2tran}
\end{equation}
where
\begin{equation}
O_{mi}^{(J)}=\langle m||O_{J}||i\rangle
\equiv
\langle j_{m} l_{m} || O_{J} || j_{i} l_{i} \rangle,
\end{equation}
\begin{equation}
\begin{split}
&T_{mi}^{(J)}=
(-1)^{j_{m}+j_{i}+J}
\sum_{n}
\left\{
\begin{array}{ccc}
j_{i} & j_{m} & J\\
L & L & j_{n}
\end{array}
\right\}
\langle m||O_{L}||n\rangle \langle n||P_{L}||i\rangle\\
&+(-1)^{j_{m}+j_{i}+1}\sum_{j}
\left\{
\begin{array}{ccc}
j_{i} & j_{m} & J\\
L & L & j_{j}
\end{array}
\right\}
\langle m||O_{L}||j\rangle \langle j||P_{L}||i\rangle,
\end{split}
\end{equation}
and
\begin{equation}
\begin{split}
&T_{mnij}^{(J_{p}J_{h}J)}=
2\frac{\hat{J}_{p}}{\sqrt{1+\delta_{mn})}}\frac{\hat{J}_{h}}{\sqrt{1+\delta_{ij}}}
\left[\left(
\begin{array}{ccc}
j_{m} & j_{n} & J_{p}\\
j_{i} & j_{j} & J_{h}\\
L & L & J\\
\end{array}
\right)\right.\\
&\left.\times O^{(L)}_{mi}P^{(L)}_{nj}
-(-1)^{j_{i}+j_{j}-J_{h}}
\left(
\begin{array}{ccc}
j_{m} & j_{n} & J_{p}\\
j_{j} & j_{i} & J_{h}\\
L & L & J\\
\end{array}
\right)
O_{mj}^{(L)} P^{(L)}_{ni} \right].
\end{split}
\end{equation}
The first term of Eq.~\eqref{eq:2tran} represents the transition strength for 1p1h excitation and appears only when the external field is separable and local. 
Although it is irrelevant to the 2-body external field like the meson exchange currents, we discuss the transition strength including it.
We also define the RPA amplitudes, which indicate how extent 1p1h and 2p2h states contribute to form a certain resonance, defined by
\begin{equation}
    P_{1p1h}^{(\nu)}=\sum_{mi}\left(X_{mi}^{(\nu)}\right)^{2}-\left(Y_{mi}^{(\nu)}\right)^{2}=\sum_{mi}p_{mi}^{(\nu)}
    \label{eq:amp1}
\end{equation}
and
\begin{equation}
    P_{2p2h}^{(\nu)}=\sum_{m<n,i<j}\left(\mathcal{X}_{mnij}^{(\nu)}\right)^{2}-\left(\mathcal{Y}_{mnij}^{(\nu)}\right)^{2}=\sum_{m<n, i<j}p_{mnij}^{(\nu)}.
    \label{eq:amp2}
\end{equation}
\par
The normalization condition $P_{1p1h}^{(\nu)}+P_{2p2h}^{(\nu)}=1$ holds for resonance state $\nu$. 
The amplitudes $p_{mi}^{(\nu)}$ and $p_{mnij}^{(\nu)}$ give us information on how extent 1p1h ($mi$) and 2p2h ($mnij$) states contribute to a resonance state $\nu$, respectively.
We here call $P_{mi}^{(\nu)}$ and $P_{mnij}^{(\nu)}$ as total 1p1h and 2p2h amplitudes, respectively.
\par
It is known that strength distributions calculated by SRPA are significantly smaller than 1p1h RPA, because some part of matrix elements of the two-body force are already taken into account in the Skyrme-Hartree-Fock~\cite{Tselyaev2007}.
To overcome this problem, the subtraction method~\cite{Gambacurta2015} is applied.
We do not go into the detail in this paper because the space is limited, but the validity is discussed in some papers~\cite{Gambacurta2015,Yang2021}.
The SRPA calculation is performed with energies of the single-particle and 2p2h excitation energies less than $70$ MeV.
The diagonal approximation that neglects transitions between 2p2h states is adopted.
\begin{figure}
\centering
\includegraphics[width=0.9\linewidth]{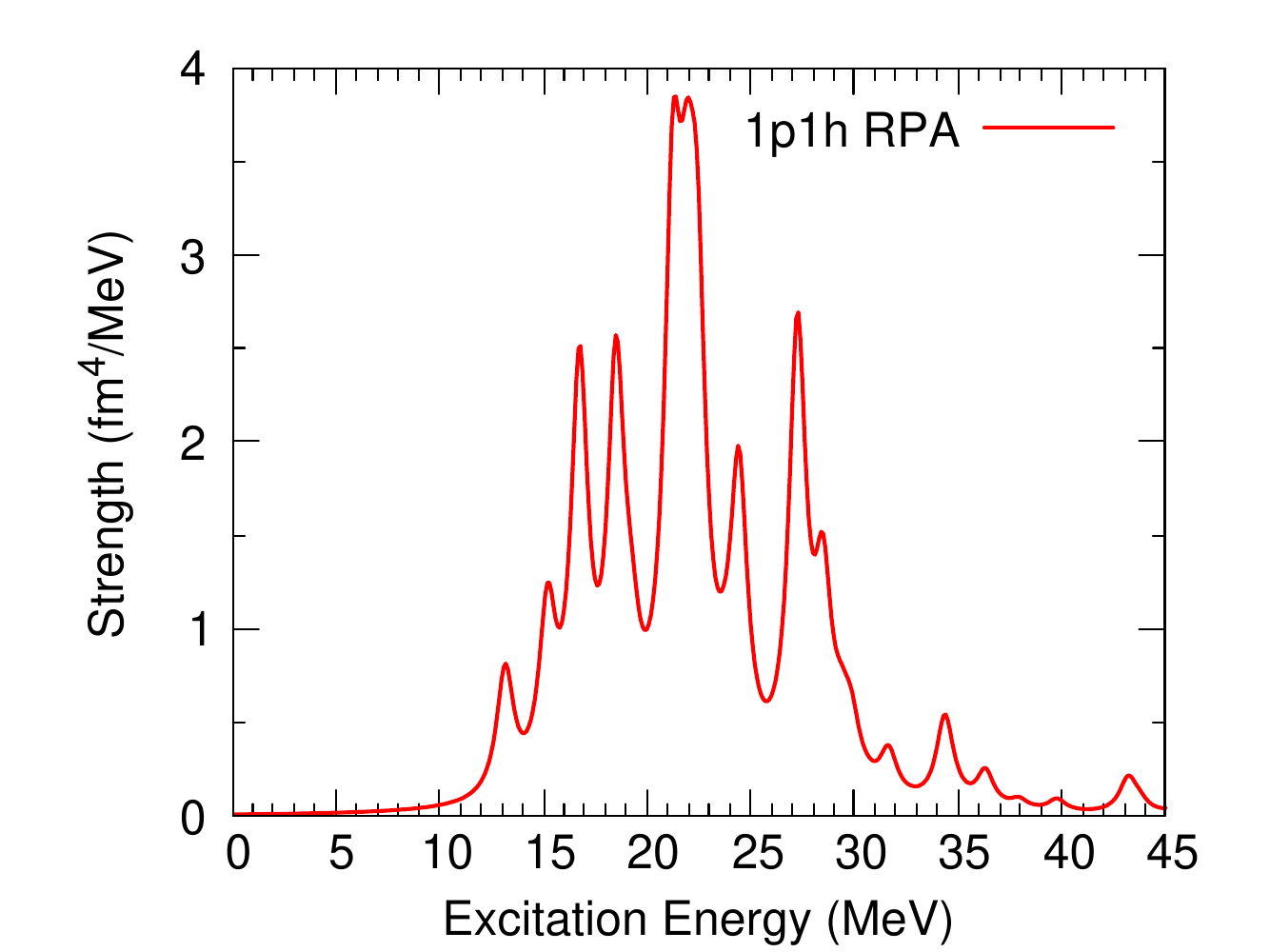}
\caption{Calculated strength distribution of isoscalar quadrupole excitations for $^{16}$O. The strength distribution is smoothed by a Lorentzian function with a width $1$ MeV.}
\label{fig:Quadru}
\end{figure}
\begin{figure}
\centering
\includegraphics[width=0.9\linewidth]{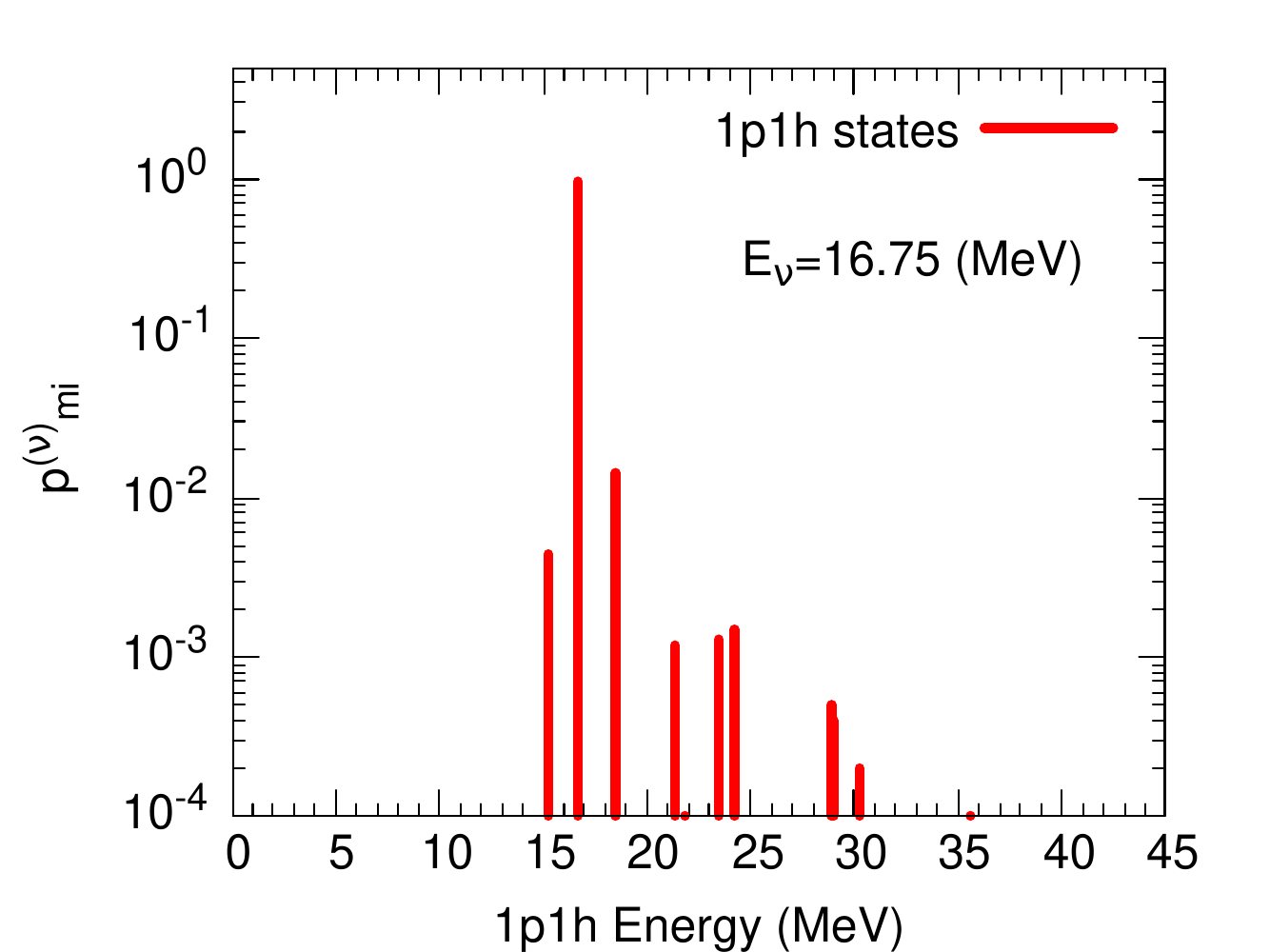}
\caption{RPA amplitude, $p_{1p1h}^{(\nu)}$, for the GQR at $E_{\nu}=16.75$ MeV for $^{16}$O.}
\label{fig:AmpQ}
\end{figure}
\section{Result}
\label{sect:result}
First of all, we discuss ordinal nuclear 1p1h excitations before going to discussion on 2p2h excitations.
Since computational time is huge for higher angular momentum of resonance, we mainly discuss isoscalar monopole resonances, namely, $O_{L}=r^{2}Y_{0}(\hat{r})$.
Figure~\ref{fig:Quadru} show the result of isoscalar monopole excitation for $^{16}$O.
The strength distribution is smoothed by a Lorentzian function with a width $1$ MeV.
The monopole resonances are observed around $E_{\nu}=22.00$ MeV.
Here, we focus on RPA amplitude of the monopole resonance at $E_{\nu}=16.75$ MeV.
Figure~\ref{fig:AmpQ} shows the RPA amplitude defined in Eq.~\eqref{eq:amp1} for the monopole resonance.
We can observe only a few 1p1h states around $E=15$ MeV meaningfully contribute to the resonance.
The most important 1p1h state ($p_{mi}^{(\nu)}\simeq0.97$) is proton $0p_{1/2}\rightarrow 2p_{5/2}$ at $16.71$ MeV.
We will show how these configurations change if the correlation of $2p2h$ states are included in nuclear excitations with the framework of SRPA.
\par
\begin{figure}
\centering
\includegraphics[width=0.90\linewidth]{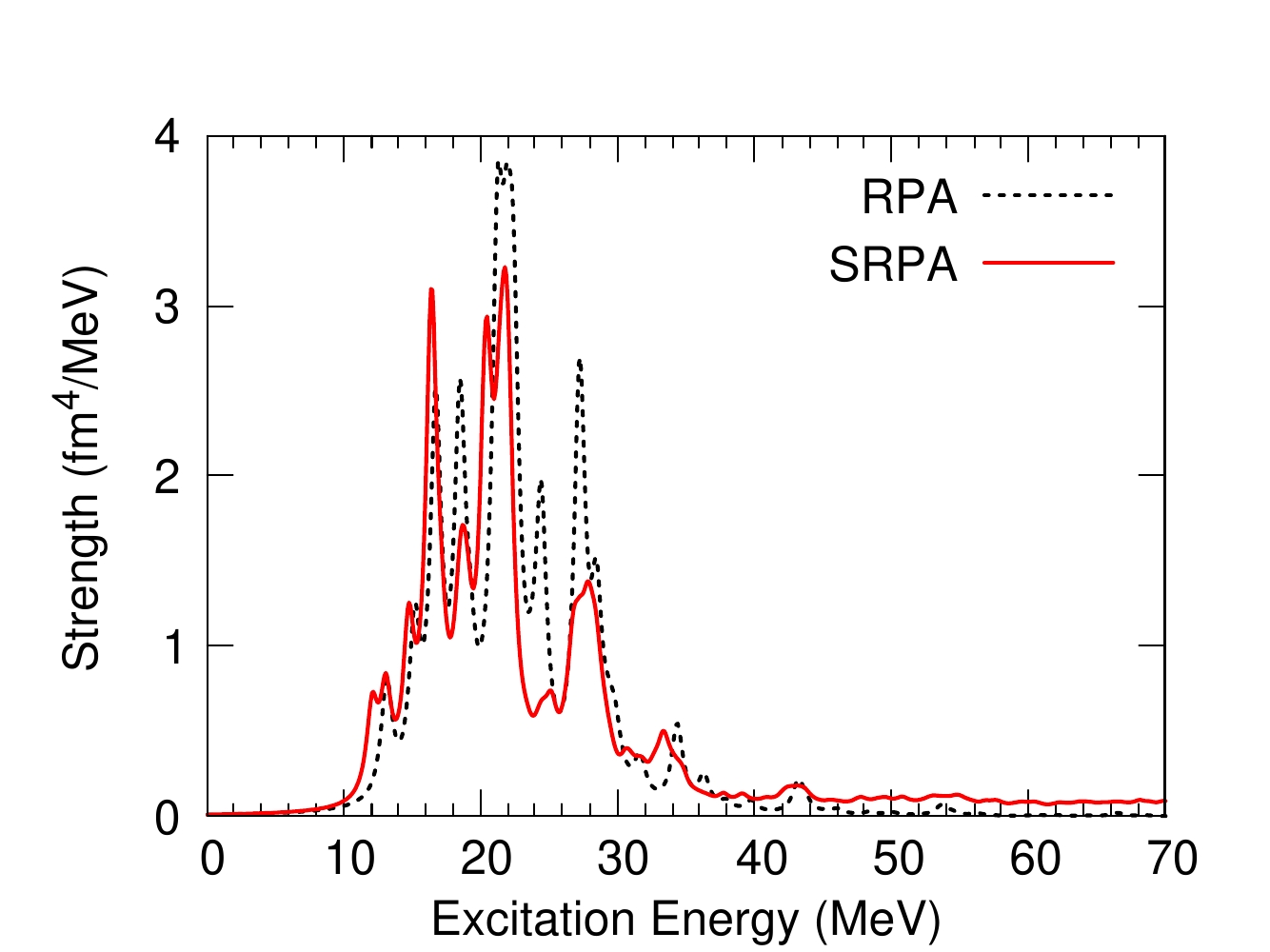}
\caption{Strength distribution of isoscalar monopole excitations for $^{16}$O. The dotted and solid lines are the results for RPA and SRPA, respectively. The lines are smoothed by a Lorentzian function with a width $1$ MeV.}
\label{fig:Monopole}
\end{figure}
\begin{figure}
\centering
\includegraphics[width=0.9\linewidth]{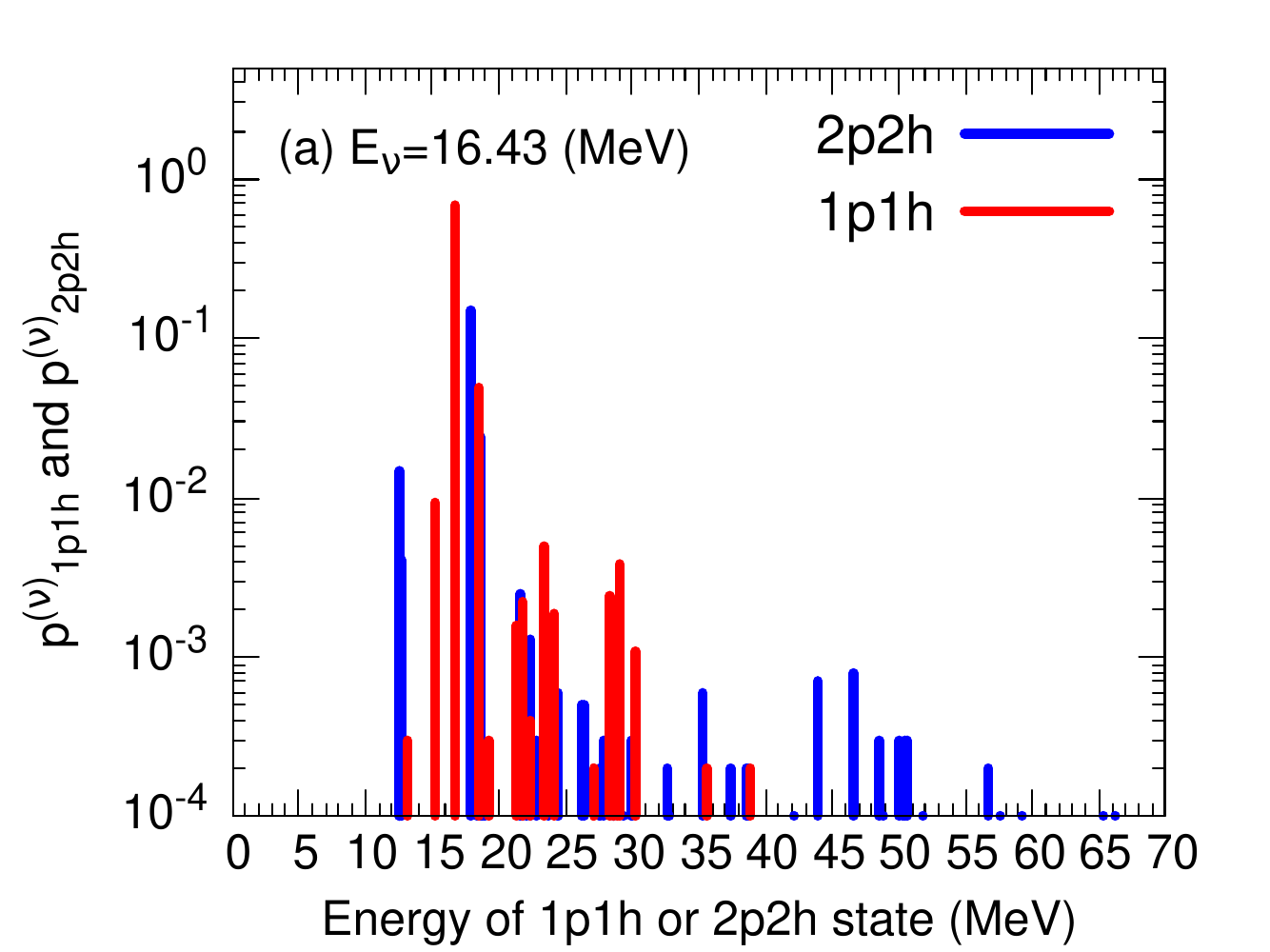}
\includegraphics[width=0.9\linewidth]{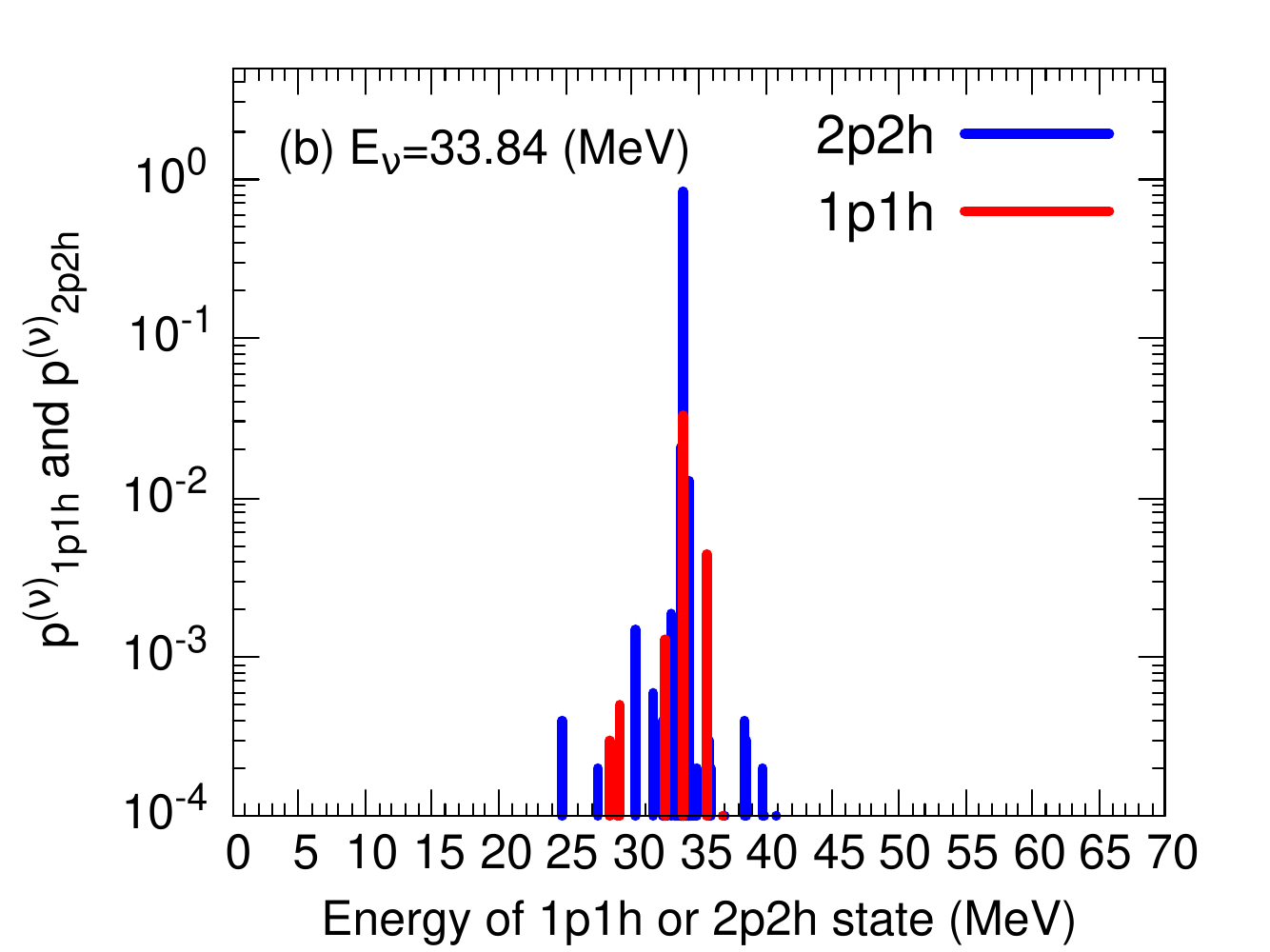}
\caption{RPA amplitudes for $^{16}$O. (a) Monopole resonance at $E_{\nu}=16.43$ MeV (b) $E_{\nu}=33.84$ MeV.}
\label{fig:MonopoleP}
\end{figure}

Figure~\ref{fig:Monopole} compares the monopole strength distributions for $^{16}$O obtained by RPA and SRPA.
The distributions of RPA and SRPA looks similar, but there are significant strengths in high energy above $45$ MeV in case of SRPA that are resulted from the coupling with 2p2h states.
We now pay attention to RPA amplitudes of 2 resonances of SRPA, which are $E_{\nu}=16.43$ MeV and $33.84$ MeV.
The RPA amplitudes of those resonances are shown in Fig.~\ref{fig:MonopoleP}.
For the low-lying resonance at $E_{\nu}=16.43$ MeV, the most important 1p1h state remain the same as RPA (proton $0p_{1/2}\rightarrow 2p_{5/2}$), but the RPA amplitude decreases to $p_{mi}^{(\nu)}\simeq0.69$. 
In turn, other 1p1h states, especially neutron $0p_{1/2}\rightarrow 2p_{5/2}$ ($p_{mi}^{(\nu)}\simeq0.05$), become significant. 
Moreover, 2p2h states come to contribute in a comparable level to 1p1h states.
For the high-lying resonance at $E_{\nu}=33.84$ MeV, the contribution of 2p2h states becomes larger than 1p1h states.
The most important configuration is neutron $0p_{1/2}\rightarrow1s_{1/2}$ and proton $0p_{3/2}\rightarrow2d_{5/2}$ with $p_{mnij}^{(\nu)}=0.83$.
This fact is reasonable because the number of possible configurations that 2p2h can take increase with the excitation energy.
If this resonance state is generated in some way, 1p1h states are thus easy to transit to 2p2h states.
We also notice that the resonance at $E_{\nu}=16.43$ MeV have various 1p1h as well as 2p2h states ranging low to higher energies contribute, while that at $E_{\nu}=33.43$ MeV have contributions of 1p1h and 2p2h states in a limited energy region from $25$ to $40$ MeV, losing the nuclear collectivity.
\par
\begin{figure}
\centering
\includegraphics[width=0.9\linewidth]{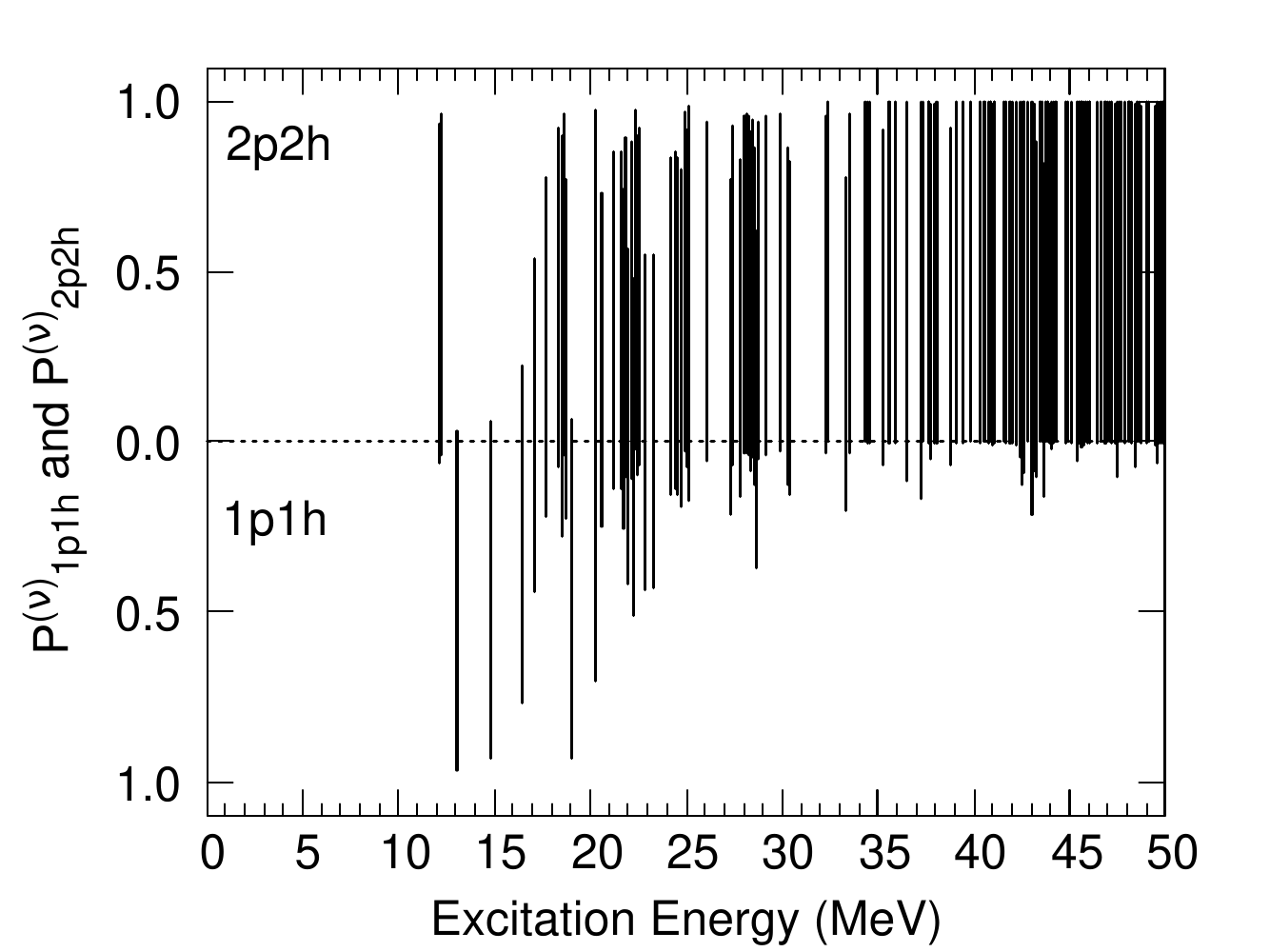}
\caption{$P_{1p1h}^{(\nu)}$ and $P_{2p2h}^{(\nu)}$ as a function of excitation energy, $E_{\nu}$. Only the resonances with the transition strengths $>0.5$ fm$^{2}$ are shown.}
\label{fig:amplitudes}
\end{figure}

Total 1p1h and 2p2h amplitudes ($P_{1p1h}^{(\nu)}$ and $P_{2p2h}^{(\nu)}$) calculated by SRPA as a function of excitation energy are shown in Fig.~\ref{fig:amplitudes}.
We only show the resonances with the transition strengths $>0.5$ fm$^{4}$.
There are resonances around $E_{\nu}=12.0$~MeV which 2p2h state predominantly contributes.
One of them is the first $0^{+}$ state that cannot be produced by the ordinal 1p1h RPA as already discussed in Ref.~\cite{Yang2021}.
Except the resonances, 1p1h states are major parts of resonances at $12.0<E_{\nu}<17.0$ MeV.
However, from $E_{\nu}>17$ MeV, the contributions of 2p2h states again become significant and those of 1p1h states changes to a minor contribution.
\par
\begin{figure}
\centering
\includegraphics[width=0.9\linewidth]{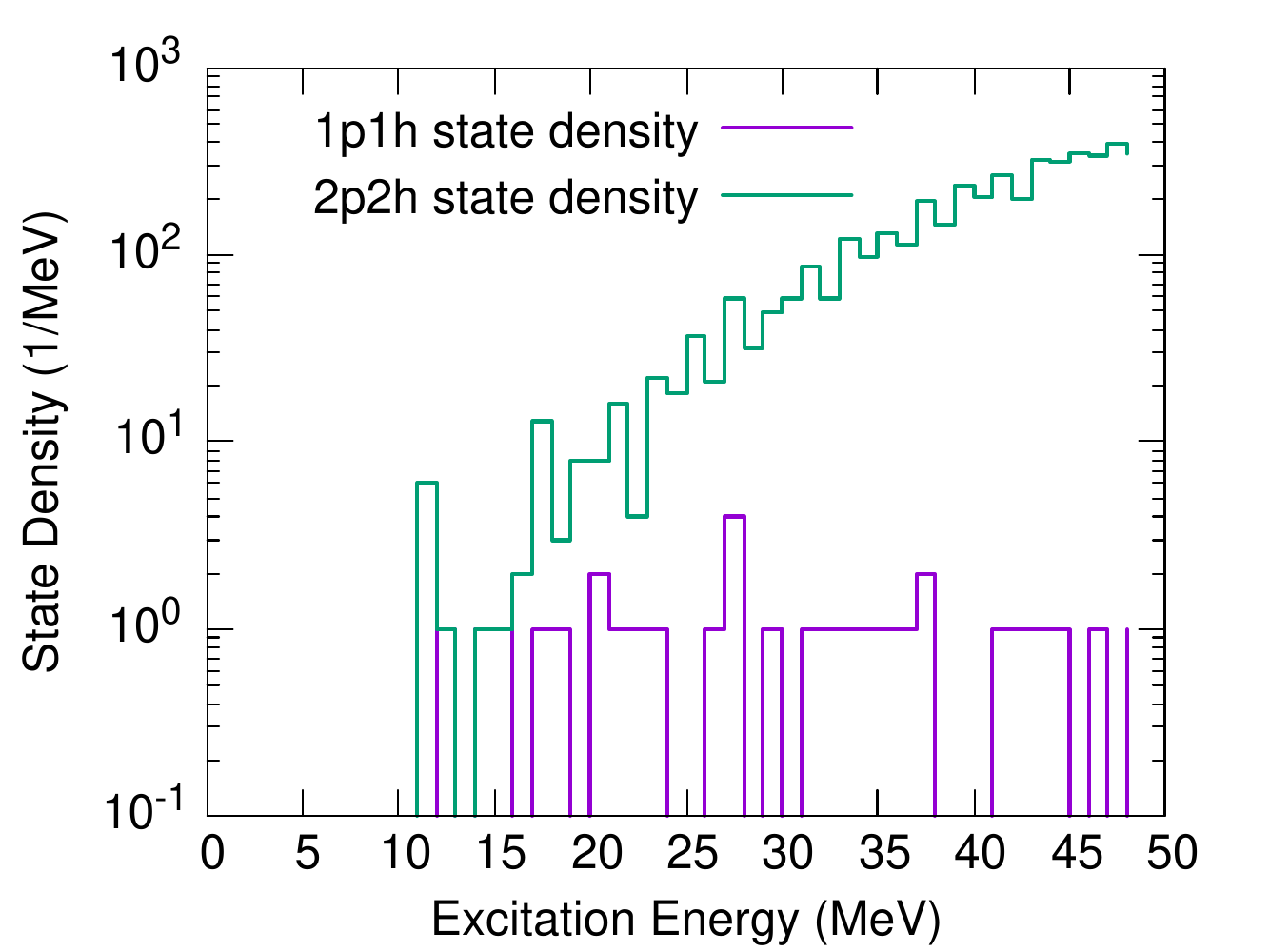}
\caption{1p1h and 2p2h state densities of $J^{\pi}=0^{+}$ for $^{16}$O.}
\label{fig:statedensity}
\end{figure}
\begin{figure}
\centering
\includegraphics[width=0.9\linewidth]{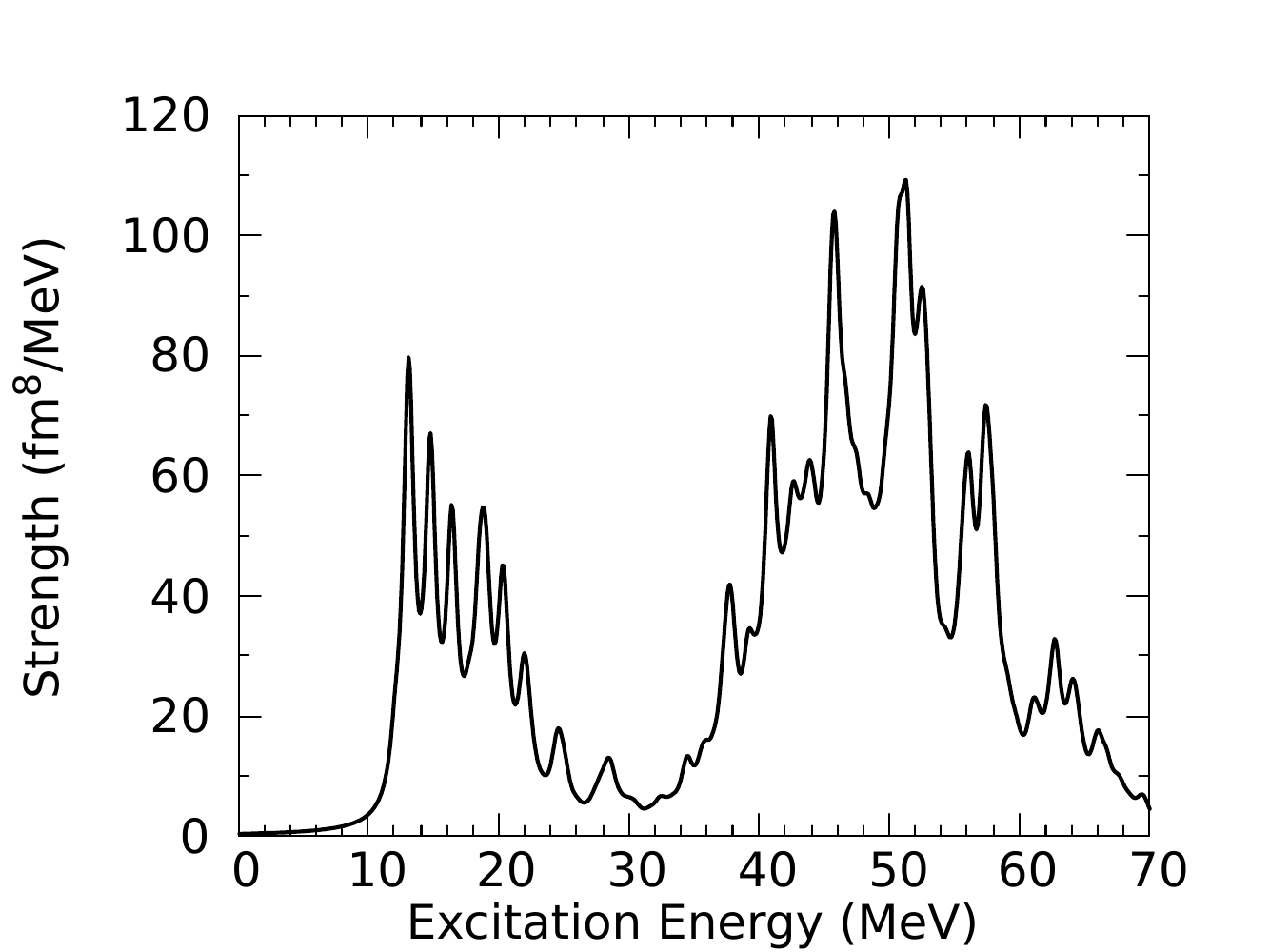}
\caption{Calculated strength distribution of double isoscalar monopole operator for $^{16}$O. The strength distribution is smoothed by a Lorentzian function with a width 1 MeV.}
\label{fig:doublemono}
\end{figure}

To study further the variations of RPA amplitudes with the excitation energy, we illustrate the 1p1h and 2p2h state densities of $J^{\pi}=0^{+}$ for $^{16}$O in Fig.~\ref{fig:statedensity}.
The 1p1h state density is almost constant with the excitation energy, while the 2p2h state density exponentially increases.
The 2p2h level density exceeds 1p1h one at about $15$ MeV.
This is consistent to the result of Fig.~\ref{fig:amplitudes}, where the 2p2h state becomes significant from $E_{\nu}=17$ MeV.
Next, we discuss the result of strength distribution generated by the 2-body excitation.
The strength distribution of double isoscalar monopole operator $T_{J}=[r^{2}Y_{0} \cdot r^{2} Y_{0}]^{0}$ is shown in Fig.~\ref{fig:doublemono}.
There are two groups of resonance around $E_{\nu}=18$ and $50$ MeV.
The resonances of the low-lying and high-lying groups are attributed from the first and second terms of Eq.~\eqref{eq:2tran}, respectively.
As mentioned above, the first term is however irrelevant to 2-body excitations. 
\par
\begin{figure}
\includegraphics[width=0.9\linewidth]{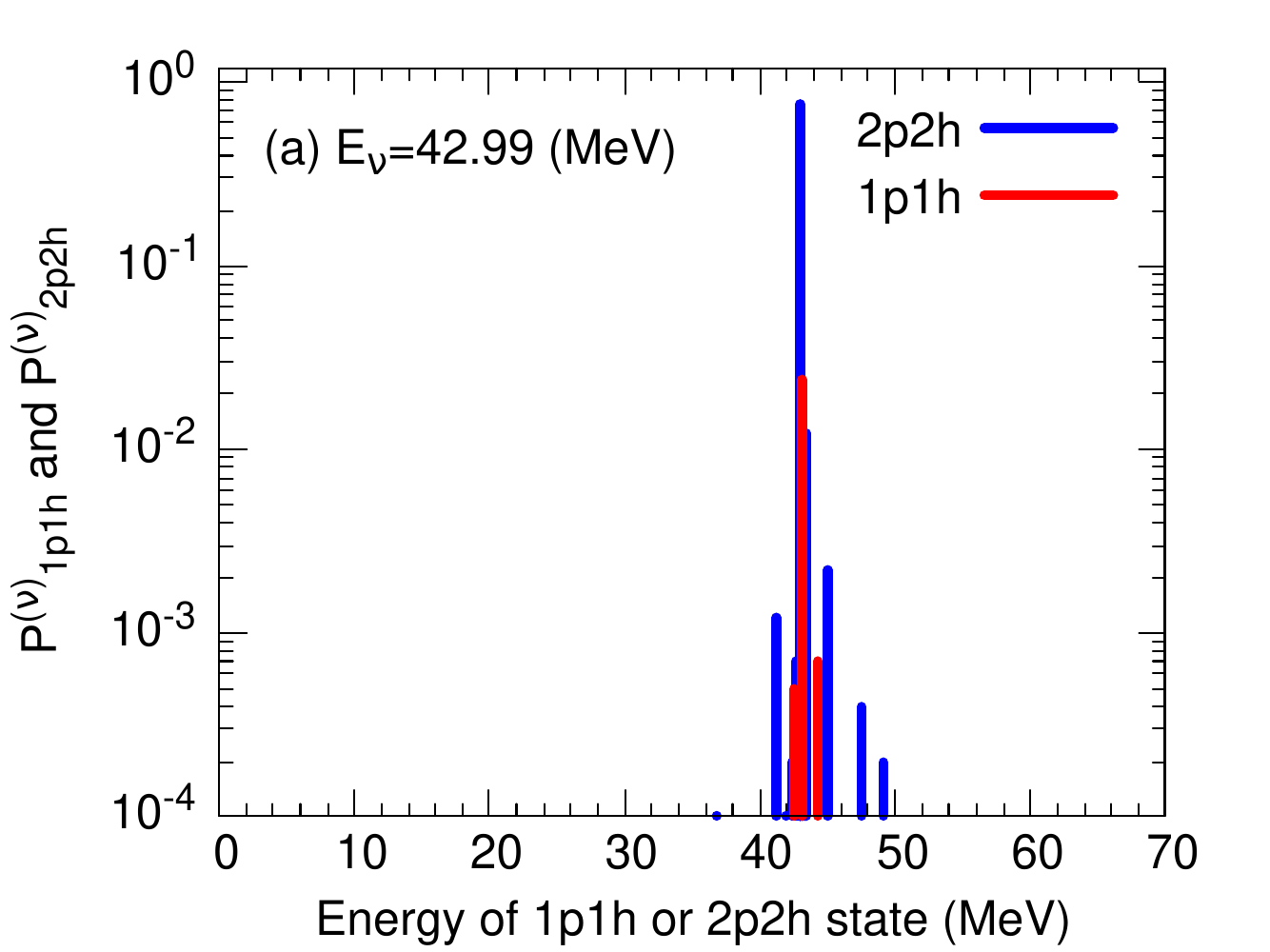}
\includegraphics[width=0.9\linewidth]{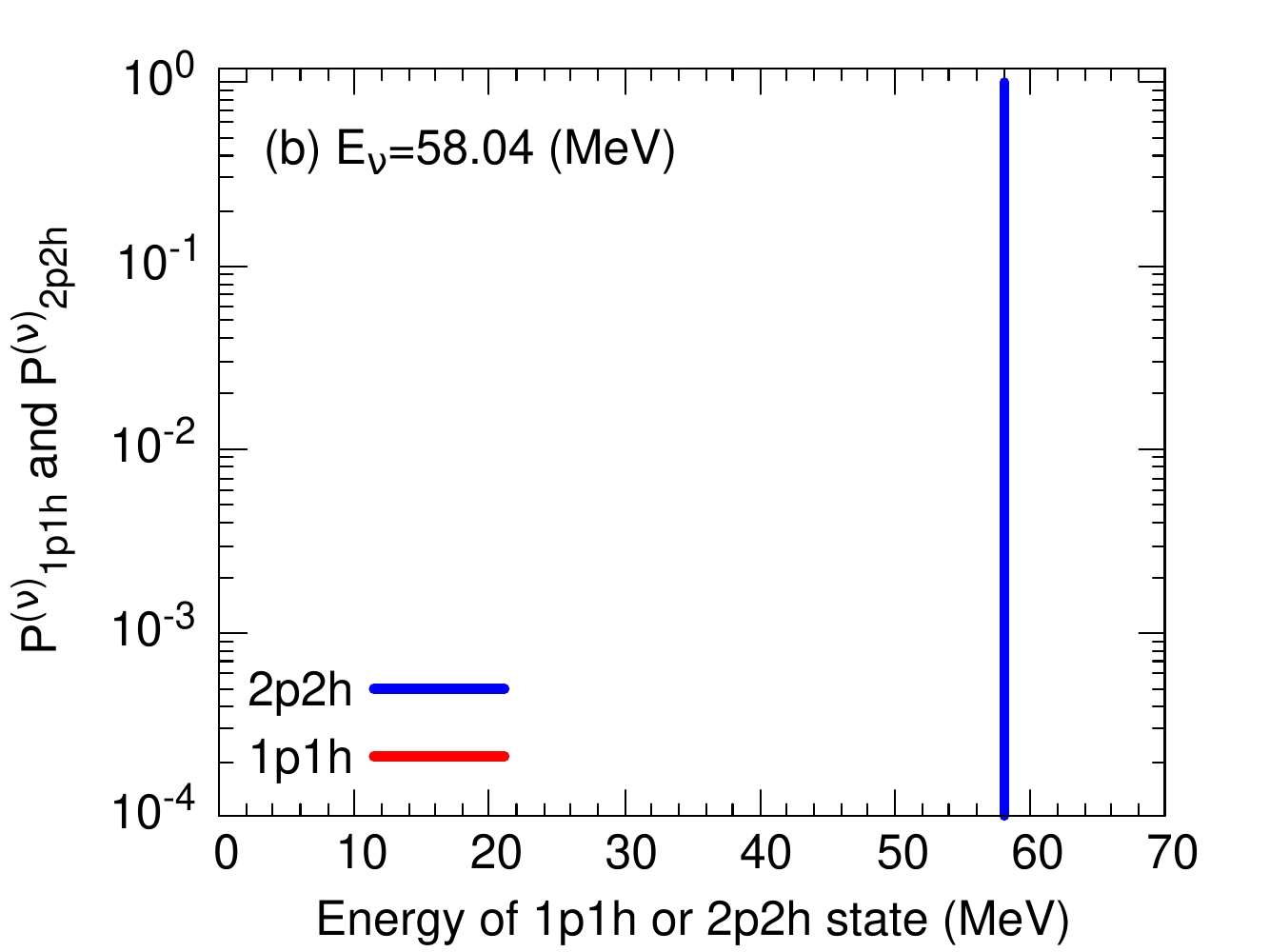}
\caption{RPA amplitudes of double isoscalar monopole resonances at (a) $E_{\nu}=42.99$ MeV (b) $E_{\nu}=58.04$ MeV for $^{16}$O.}
\label{fig:amplitudes2}
\end{figure}

We picked up 2 resonances at $E_{\nu}=42.99$ and $58.04$ MeV from the high-lying resonance group.
The RPA amplitudes of them are shown in Fig.~\ref{fig:amplitudes2}.
For $E_{\nu}=49.99$ MeV, the the main configuration is proton ($(0p_{1/2})^{2}\rightarrow0d_{5/2}3d_{5/2}$ with $p_{mnij}^{(\nu)}=0.72$.
The 1p1h and 2p2h states contributing the resonance distribute in a very limited region as compared to the other resonances shown in Fig.~\ref{fig:MonopoleP}, indicating the nuclear collectivity becomes weak at this energy.
For $E=58.4$ MeV, only one peak coming from 2p2h states contribute the resonances.
The main configuration is neutron $0p_{3/2}\rightarrow2p_{3/2}$ and proton $0p_{3/2}\rightarrow 4p_{3/2}$.
Nuclear collectivity is meaningfully lost at high excitation energies where double isoscalar monopole resonances are significant.

\section{Summary}
\label{sect:summary}
We studied how 1p1h states generated by 1-body external fields transit to 2p2h states within SRPA.
We found that 2p2h states contribute resonances significantly with increasing excitation energy.
It is accordingly expected that initial 1p1h states created by a 1-body external field transit to 2p2h states rapidly.
Even for $^{16}$O, the effect of 2p2h states cannot be neglected from a low excitation energy.
\par
The RPA amplitudes of resonances generated by 2-body external fields were also discussed.
We found that at energies where double isoscalar monopole excitations are found the resonances were mainly created by 2p2h states.
Nuclear collectivity rapidly decreases with increasing excitation energy.
This outcome indicates that the independent particle picture may be valid when discussing the effect of MEC in nuclei.
We did not consider transition between 2p2h states because we adopted diagonal approximation.
We plan to study further how the present result changes when the 2p2h residual interactions are fully included, not using the diagonal approximation.
Our perspective is to apply the SRPA method to study the effect of MEC in muon captures.
New experiments on particle emissions following muon capture are planned and the results will help us confirm the validity of SRPA.
%
\paragraph{Acknowledgement}
This work was supported by JSPS KAKENHI Grant Numbers JP23K03426 and JP24K00647.
\bibliography{ref}
\end{document}